\documentclass[12pt]{article}
\usepackage{enumerate}
\usepackage{amsfonts}
\usepackage{amsmath}
\usepackage{amssymb}
\usepackage{amsthm}

\def\H{\mathcal{H}}

\def\S{\mathfrak{S}}

\def\T{\mathfrak{T}}
\def\B{\mathfrak{B}}

\newcommand{\id}{\mathrm{Id}}
\newcommand{\Tr}{\mathrm{Tr}}

\newcommand{\shs}{\hspace{1pt}}

\newcounter{defin}  \newcounter{lemma}  \newcounter{theorem}
\newcounter{property} \newcounter{corol}  \newcounter{remark} \newcounter{example}

\newenvironment{lemma}{\par\refstepcounter{lemma}
     \textbf{Lemma \thelemma.} }{\rm\par}

\newenvironment{property}{\par\refstepcounter{property}
     \textbf{Proposition \theproperty.}\ }{\rm\par}

\newenvironment{definition}{\par\refstepcounter{defin}
     \textbf{Definition \thedefin.}\ }{\rm\par}
\newenvironment{remark}{\par\refstepcounter{remark}
     \textbf{Remark \theremark.}}{\rm\par}

\begin{document}

\title{Lower bounds on distances between a given quantum channel and certain classes of channels}
\author{M.E. Shirokov\footnote{Steklov Mathematical Institute, RAS, Moscow, email:msh@mi.ras.ru} , A.V.Bulinski\footnote{Moscow Institute of Physics and Technology, email:andrey.bulinski@yandex.ru
}}
\date{}
\maketitle

\begin{abstract}
The tight, in a sense,
lower estimates of diamond-norm distance from a given quantum channel to the sets of degradable, antidegradable and entanglement-breaking channels are obtained via the tight continuity bounds
for quantum mutual information and for relative entropy of entanglement in finite-dimensional case.
As an auxiliary result there are established
lower bounds of trace-norm distance from a given bipartite state to the set of all separable states.
\end{abstract}

\section{Introduction and basic notions}

For a continuous numerical function on a topological space one often deals with assessing the variability of increments by means of continuity bounds, seldom named variation bounds when no ambiguity with the term variation  arises.
The bounds on increments of entropic characteristics of quantum states and channels are quite useful to analyze the issues where the uniform continuity of those characteristics is important. Just recall the well-known Fannes and Alicki-Fannes continuity bounds for von Neumann entropy and for quantum conditional entropy, respectively, which play the key role in deriving a number of
quantum information theory results \cite{A&F,Fannes,H-SCI,L&S,Wilde}.

In this paper, it is demonstrated that one can also employ such continuity bounds to establish the lower bounds on distance from a given quantum state (channel) to a certain class of states (channels). In other words, the mentioned bounds on variability enable one to estimate, for a given state (channel), the size of its neighborhood devoid of the states (channels) of certain type. Exactly in this nonstandard application of continuity bounds their accuracy is essential.

The paper is organized as follows: the rest of introductory Sec.1 fixes some notation and terminology; in Sec.2 the method of nonstandard applying continuity bounds is presented;
in Sec.3 the lower bounds on distance between a bipartite state and the set of separable states are deduced. Sec.4 contains the main results for channels and illustrative examples.
\medskip

Let $\mathcal{H}$ be a finite-dimensional
Hilbert space,
$\mathfrak{B}(\mathcal{H})$ the algebra of all (linear) operators in $\mathcal{H}$ endowed with the operator norm $\|\cdot\|$ and $\mathfrak{T}( \mathcal{H})$ the
Banach space of all operators   with the trace norm $\|\!\cdot\!\|_1$.
Denote by $\mathfrak{T}_{+}(\mathcal{H})$ the cone of positive operators in $\mathfrak{T}(\mathcal{H})$  and by $\mathfrak{S}(\mathcal{H})$ the convex set of density operators, i.e. elements of $\mathfrak{T}_{+}(\mathcal{H})$ with unit trace, describing the \textit{quantum states} \cite{H-SCI,Wilde}.

Let $I_{\mathcal{H}}$ be the unit operator in a Hilbert space
$\mathcal{H}$ and $\mathrm{Id}_{\mathcal{H}}$ the identity
transformation of the Banach space $\mathfrak{T}(\mathcal{H})$.\smallskip

The \emph{von Neumann entropy}  of a quantum
state $\rho\in\mathfrak{S}(\mathcal{H})$ defined by the formula $H(\rho)=\mathrm{Tr}\eta(\rho)$, where $\eta(x)=-x\log x$ for $x>0$ and $\eta(0)=0$,
is a nonnegative concave continuous function on $\mathfrak{S}(\mathcal{H})$ \cite{H-SCI,Wilde}. Below we will use the binary entropy $h_2(x)=\eta(x)+\eta(1-x)$ as well.

The \emph{quantum relative entropy} for two states $\rho$ and
$\sigma$ in $\mathfrak{S}(\mathcal{H})$ is defined (see \cite{H-SCI,Wilde}) by way of
$$
H(\rho\shs\|\shs\sigma)=\Tr(\rho\log\rho-\rho\log\sigma),
$$
whenever $\mathrm{supp}\,\rho \subseteq \mathrm{supp}\hspace{1pt}\sigma $, and $H(\rho \,\Vert \sigma )=+\infty $, otherwise.
Here  $\mathrm{supp}\,\rho$
stands for the orthogonal complement of the zero eigenspace $\mathrm{ker}\,\rho$.

If quantum systems $A$ and $B$ are described by Hilbert spaces  $\H_A$ and $\H_B$ then the bipartite system $AB$ is described by the tensor product of these spaces, i.e. $\H_{AB}\doteq\H_A\otimes\H_B$. A state in $\S(\H_{AB})$ is denoted $\rho_{AB}$, its marginal states $\Tr_{B}\rho_{AB}$ and $\Tr_{A}\rho_{AB}$ are denoted $\rho_{A}$ and $\rho_{B}$ respectively where $\Tr_{B}$
is the partial trace over $\H_B$ (and similarly for $A$), see, e.g.,\cite{H-SCI}.

\emph{Quantum mutual information} $\,$ of a composite quantum system in the state $\,\omega _{AB}$ is defined by the following expressions (see \cite{L-mi})
\begin{equation*}
I(A\!:\!B)_{\omega }=H(\omega _{AB}\hspace{1pt}\Vert \hspace{1pt}\omega
_{A}\otimes \omega _{B})=H(\omega _{A})+H(\omega _{B})-H(\omega _{AB}).
\end{equation*}

By invoking the optimization of Alicki-Fannes method due to Winter \cite{W-CB} it was shown in \cite{CHI} that, for any states $\rho$ and $\sigma$ in $\,\S(\H_{AB})$, one has
\begin{equation}\label{MI-CB}
|I(A\!:\!B)_{\rho}-I(A\!:\!B)_{\sigma}|\leq 2\varepsilon
\log d+2g(\varepsilon)
\end{equation}
where $\;\varepsilon=\frac{1}{2}\|\shs\rho-\sigma\|_1\,$, $\,d=\min\{\dim\H_A,\dim\H_B\}$,  $\,g(\varepsilon)\!=\!(1+\varepsilon)h_2\!\left(\frac{\varepsilon}{1+\varepsilon}\right)$, and that the term $2g(\varepsilon)$ in this inequality can be replaced by $g(\varepsilon)$ if either
$\rho_A=\sigma_A$ or $\rho_B=\sigma_B$.
We call (\ref{MI-CB}) a \emph{continuity bound} for quantum mutual information on $\,\S(\H_{AB})$.  It is tight for large $d$ in the sense of the following
\smallskip
\begin{definition}\label{tight-cb}
A continuity bound $\,\displaystyle|F(x)-F(y)|\leq B_a(x,y),\;x,y\in S_a,\,$ depending on a parameter $\,a\,$ is called \emph{tight} for large $\,a\,$ if $$\;\displaystyle\limsup_{a\rightarrow+\infty}\sup_{x,y\in S_a}\frac{|F(x)-F(y)|}{B_a(x,y)}=1.$$
\end{definition}

We will also use the following
\smallskip
\begin{definition}\label{tight-lb}
A lower  bound $\,\displaystyle F(x)\geq B_a(x),\;x\in S_a,\,$ for a nonnegative function $F$ depending on a parameter $\,a\,$ is called \emph{tight} for large $\,a\,$ if $$\;\displaystyle\limsup_{a\rightarrow+\infty}\sup_{x\in S_a}\frac{B_a(x)}{F(x)}=1.$$
\end{definition}

A \emph{quantum channel} $\,\Phi$ from a system $A$ to a system
$B$ is a completely positive trace preserving linear map
$\mathfrak{T}(\mathcal{H}_A)\rightarrow\mathfrak{T}(\mathcal{H}_B)$,
where $\mathcal{H}_A$ and $\mathcal{H}_B$ are the respective Hilbert spaces
associated with these systems \cite{H-SCI,N&Ch,Wilde}. For brevity sake we will write $\,\Phi: A\rightarrow B$. \smallskip

Denote by $\mathfrak{F}(A,B)$ the set of all quantum channels from a system $A$ to a system
$B$. We will use two metrics  on the set $\mathfrak{F}(A,B)$ induced respectively by the operator norm
$$
\|\Phi\|\doteq \sup_{\rho\in\T(\H_A),\|\rho\|_1=1}\|\Phi(\rho)\|_1
$$
and by the diamond norm
$$
\|\Phi\|_{\diamond}\doteq \sup_{\rho\in\T(\H_{AR}),\|\rho\|_1=1}\|\Phi\otimes \id_R(\rho)\|_1
$$
of a map $\Phi:\T(\H_A)\rightarrow\T(\H_B)$. The latter  coincides with the norm of complete boundedness of the dual to $\Phi$ map $\Phi^*:\B(\H_B)\rightarrow\B(\H_A)$ \cite{H-SCI,Wilde}. \smallskip

\section{On the ``inverse" employment of continuity bounds }

Now consider a function $F : X \to \mathbb{R}$ defined on a metric space $(X,D)$. Given a  continuity bound of the form $|F(x_1)-F(x_2)|\leq f(D(x_1,x_2))$ with some ``good" $f : \mathbb{R}_{+} \to \mathbb{R}$, we are interested in a lower estimate for $D(x_1,x_2)$ in terms of $f$ and the increment of $F$. More precisely, consider the following special form of $f$.
\smallskip

\begin{lemma}\label{app}
A) \emph{Let $X$ be a set with a metric\footnote{The assertions of the lemma are valid for any nonnegative function $D$ on $X\times X$.}
$D$ and $F$ be a function on $X$ such that
$$
|F(x_1)-F(x_2)|\leq A\varepsilon+r(\varepsilon),\quad \varepsilon=D(x_1,x_2),
$$
for all $\,x_1,x_2\in X$, where $A>0$ and  $\,r$ is
a nondecreasing function on $\,\mathbb{R}_+$ such that $r(0)=0$ (occasionally we also set $r(t)=0$ for $t<0$). Then}
$$
D(x_1,x_2)\geq A^{-1}\left(\Delta-r(A^{-1}\Delta)\right),\quad \Delta=|F(x_1)-F(x_2)|.
$$
B) \emph{If $\,F(x)\leq 0$ for any $\,x\in X_0\subset X$ and $\,F(x_*)>0$ for some $x_*\in X$ then
$$
\inf_{x\in X_0} D(x_*,x)\geq A^{-1}\left(G-r(A^{-1}G)\right)
$$
for any  positive $G\leq F(x_*)$.}
\end{lemma}\medskip

\emph{Proof.} A) Since $h(\varepsilon)=\varepsilon+A^{-1}r(\varepsilon)$ is an increasing nonnegative function such that $h(0)=0$, we introduce the inverse function $h^{-1}$  for $h$ and see that

$$
\varepsilon\geq h^{-1}(A^{-1}\Delta)=A^{-1}\Delta-[A^{-1}\Delta-h^{-1}(A^{-1}\Delta)]\geq A^{-1}\Delta-A^{-1}r(A^{-1}\Delta),
$$
because the last inequality follows from the $\,r$ nondecreasing:
$$
h^{-1}(t)\geq t-A^{-1}r(t)\;\Leftrightarrow\;t\geq h(t-A^{-1}r(t))=t-A^{-1}r(t)+A^{-1}r(t-A^{-1}r(t))
$$
where $t=A^{-1}\Delta$.\smallskip

B) This follows by the above argument, since for any $x\in X_0$ we have
$$
G\leq F(x_*)\leq A\varepsilon+r(\varepsilon),\quad \varepsilon=D(x_*,x).\;\square
$$

\begin{remark}\label{app}
The indicated lower bounds for $D(x_1,x_2)$ need not be nontrivial, i.e. positive. However, in the examples below they are nontrivial.
\end{remark}\smallskip

\section{Lower bounds for the distance from a fixed bipartite state to the set of all separable states}

Let $\rho_{AB}$  be an arbitrarily fixed state in $\S(\H_{AB})$. By virtue of the part B of Lemma  \ref{app} one can get a lower bound for distance from $\rho_{AB}$ to the set $\S_\mathrm{s}(\H_{AB})$ of all the separable states\footnote{The set $\S_\mathrm{s}(\H_{AB})$ is the convex hull of all product states $\rho_A\otimes\sigma_B$ in $\S(\H_{AB})$ \cite{H-SCI,Wilde}.} in $\,\S(\H_{AB})$, i.e. for the value of
\begin{equation*}
D_\mathrm{s}(\rho_{AB})=\inf_{\sigma\in\S_\mathrm{s}(\H_{AB})}\|\rho-\sigma\|_1.
\end{equation*}
To this end one can employ a continuity bound  for any indicator of entanglement on $\S(\H_{AB})$ (being a nonnegative function $E$ on $\S(\H_{AB})$ such that  $E^{-1}(0)=\S_{\mathrm{s}}(\H_{AB})$), provided that this bound is of the form treated in Lemma 1. In particular, any asymptotically continuous measure of entanglement $E$ on $\S(\H_{AB})$ could be used \cite{P&V}.

The choice of specific function $E$ for solving the present problem is determined by the following two requirements:
\begin{itemize}
  \item existence of fairly accurate continuity bound for $E$;
  \item possibility to evaluate $E(\rho_{AB})$ for any state $\rho_{AB}$ or availability of computable lower bound for $E(\rho_{AB})$.
\end{itemize}
The first of them is due to the desire to obtain a reasonable lower estimate for $D_\mathrm{s}(\rho_{AB})$, and the second stems from the need in its computability.

Among the commonly known entanglement measures on $\S(\H_{AB})$ the choice of the \emph{relative entropy of entanglement} $E_R$ for $E$  seems optimal. Recall that, for any state $\rho$ in $\S(\H_{AB})$, one has (see \cite{V&P,P&V})
\begin{equation}\label{ree-def}
  E_R(\rho)=\inf_{\sigma\in \S_\mathrm{s}(\H_{AB})}H(\rho\shs\|\sigma).
\end{equation}
Recently Winter has established in \cite{W-CB} the following tight continuity bound for $E_R$:
\begin{equation}\label{REE-CB}
|E_R(\rho)-E_R(\sigma)|\leq \varepsilon
\log d+g(\varepsilon),
\end{equation}
valid for any states $\rho$ and $\sigma$ in $\S(\H_{AB})$, where  $d=\min\{\dim\H_A,\dim\H_B\}$, $\;\varepsilon=\frac{1}{2}\|\shs\rho-\sigma\|_1$ and $g(\varepsilon)=(1+\varepsilon)h_2\!\left(\frac{\varepsilon}{1+\varepsilon}\right)$. This significantly refines the continuity bound for $E_R$ established in \cite{D}.

Likewise all the entanglement measures, the relative entropy of entanglement is hardly amenable to computation for arbitrary states. However, it has
an easily computable lower estimate \cite{P&V}:
\begin{equation}\label{ci}
I_c(\rho)\doteq\max\{I(A\rangle B)_{\rho},I(B\rangle A)_{\rho}\}=\max\{H(\rho_{A}),H(\rho_{B})\}-H(\rho),
\end{equation}
where $I(X\rangle Y)_{\rho}\doteq -H(X|Y)_{\rho}$  is the coherent information of $\rho\in\S(\H_{AB})$. \smallskip
Taking into account continuity bound (\ref{REE-CB}) and the inequality $I_c(\rho)\leq E_R(\rho)$, upon applying Lemma \ref{app} to the function $F=E_R$ we come to the following \smallskip
\begin{property}\label{DS}
\emph{Let $\rho$ be any state in $\S(\H_{AB})$. Then
\begin{equation}\label{one}
D_\mathrm{s}(\rho)\,\geq\, 2\shs\frac{E_R(\rho)}{\log d}-\frac{2}{\log d}\;g\!\left(\frac{E_R(\rho)}{\log d}\right),
\end{equation}
where $\,d=\min\{\dim\H_A,\dim\H_B\}$ and $\,g(x)=(1+x)h_2\!\left(\frac{x}{1+x}\right)$.
If $\,I_c(\rho)>0$, where $\,I_c(\rho)$ is defined by (\ref{ci}), then
\begin{equation}\label{two}
D_\mathrm{s}(\rho)\,\geq\, 2\shs\frac{I_c(\rho)}{\log d}-\frac{2}{\log d}\;g\!\left(\frac{I_c(\rho)}{\log d}\right).
\end{equation}}
\end{property}
If $\,\H_A=\H_B$ and $\rho_{*}$ is a maximally entangled pure state in $\,\S(\H_{AB})$ then $\,E_R(\rho)=I_c(\rho)=\log d\,$ and hence (\ref{one}) and (\ref{two}) give the same lower bound
\begin{equation}\label{three}
D_\mathrm{s}(\rho_{*})\,\geq \,2-\frac{2g(1)}{\log d}=2-\frac{4\log 2}{\log d},
\end{equation}
which  provides an alternative proof of the known fact that $D_\mathrm{s}(\rho_{*})$ is close to $2$ for large $d$ \cite{W-pc}. Note also that
(\ref{three}) shows that lower bounds (\ref{one}) and (\ref{two}) both are tight for large $d$ in the sense of Def.\ref{tight-lb}.
\medskip

\section{Lower bounds for the distance from a given channel to the sets of degradable, antidegradable and entanglement-breaking channels}
\smallskip
Recall that for any quantum channel
 $\,\Phi:A\rightarrow B\,$ the Stinespring theorem guarantees the existence of a Hilbert space $\mathcal{H}_{E}$ (environment) and an isometry $V:\mathcal{H}_{A}\rightarrow \mathcal{H}_{B}\otimes \mathcal{H}_{E}$ such that
\begin{equation}\label{St-rep}
\Phi (\rho )=\mathrm{Tr}_{E}V\rho V^{\ast },\quad \rho \in \mathfrak{T}(\mathcal{H}_{A}).
\end{equation}
A quantum channel $\widehat{\Phi}: A\rightarrow E$ defined by the expression
\begin{equation}\label{c-ch}
\widehat{\Phi}(\rho)=\mathrm{Tr}_{B}V\rho V^{\ast }
\end{equation}
is called \emph{complementary} to the channel $\Phi $ \cite[Ch.6]{H-SCI}.\smallskip

 A channel $\Phi:A\rightarrow B$ is called \emph{degradable} if $\,\widehat{\Phi}=\Theta\circ\Phi\,$ for some channel $\Theta:B\rightarrow E$.
 A channel $\Phi:A\rightarrow B$ is termed  \emph{antidegradable} when $\widehat{\Phi}$ is degradable \cite{R&Co}.
Denote by $\mathfrak{F}_d(A,B)$ and $\mathfrak{F}_a(A,B)$ the sets of all degradable and antidegradable channels between the systems $A$ and $B$. These sets have a nonempty intersection, for instance, the erasure channel
\begin{equation}\label{era-ch}
\Phi_p(\rho)=\left[\begin{array}{cc}
(1-p)\rho &  0 \\
0 &  p\Tr\rho
\end{array}\right], \quad p\in[0,1],
\end{equation}
 from a $d$-dimensional system $A$ into a $(d+1)$-dimensional system $B$ is simultaneously degradable and antidegradable for $p=1/2$.
 The set $\,\mathfrak{F}_a(A,B)$ contains an important subset  $\,\mathfrak{F}_{eb}(A,B)$ of  \emph{entanglement-breaking} channels, i.e. the channels $\Phi:A\rightarrow B$ such that $\,\Phi\otimes\id_R(\omega_{AR})\,$ is a separable state in $\S(\H_{BR})$ for any state $\omega_{AR}$,  $R$ being an arbitrary quantum system \cite{R&Co,H-SCI,Wilde}.

An important property of any degradable (resp. antidegradable) channel $\Phi$ consists in nonnegativity (resp. nonpositivity) of coherent information
\begin{equation*}
I_\mathrm{c}(\Phi,\rho)=H(\Phi(\rho))-H(\widehat{\Phi}(\rho))
\end{equation*}
for any input state $\rho$ \cite{H-SCI, Wilde}. Having observed that
\begin{equation*}
I_\mathrm{c}(\Phi,\rho)=I(B\!:\!R)_{\Phi\otimes\mathrm{Id}_{R}(\hat{\rho})}-H(\rho),
\end{equation*}
where $\mathcal{H}_R\cong\mathcal{H}_A$ (are isomorphic), $\hat{\rho}$ being
a pure state in $\S(\H_{AR})$ such that $\hat{\rho}_{A}=\rho$, it is not difficult to infer from  (\ref{MI-CB}) the following continuity bound for coherent information viewed as a function of channel
\begin{equation}\label{CI-CB}
|I_\mathrm{c}(\Phi,\rho)-I_\mathrm{c}(\Psi,\rho)|\leq 2\varepsilon
\log d +g(\varepsilon).
\end{equation}
Here $\Phi$ and $\Psi$ are arbitrary channels from $A$ to $B$, $\,\varepsilon=\frac{1}{2}\|\Phi-\Psi\|_{\diamond}$, $\,d=\min\{\dim\H_A,\dim\H_B\}\,$ and $\,g(\varepsilon)=(1+\varepsilon)h_2\!\left(\frac{\varepsilon}{1+\varepsilon}\right)$.\footnote{By noting that $\,I_\mathrm{c}(\Phi,\rho)=-H(R|B)_{\Phi\otimes\mathrm{Id}_{R}(\hat{\rho})}\,$ continuity bound (\ref{CI-CB}) with $d=\dim\H_A$ can be obtained from Winter's tight continuity bound for the conditional entropy \cite{W-CB}.}\medskip

Our main result provides the lower bounds for the values of
\begin{equation*}
D_d(\Phi)\doteq\inf_{\Psi\in\mathfrak{F}_d(A,B)}\|\Phi-\Psi\|_{\diamond},\qquad D_a(\Phi)\doteq\inf_{\Psi\in\mathfrak{F}_a(A,B)}\|\Phi-\Psi\|_{\diamond},
\end{equation*}
and
\begin{equation*}
D_{eb}(\Phi)\doteq\inf_{\Psi\in\mathfrak{F}_{eb}(A,B)}\|\Phi-\Psi\|_{\diamond},
\end{equation*}
which thus determine, in terms of coherent information, the radii of the maximal open ball neighborhoods of a channel $\Phi$, devoid of degradable, antidegradable and entanglement breaking channels, respectively.  \smallskip

\begin{property}\label{VLB}\emph{ Let $\,\Phi:A\rightarrow B$ be a quantum channel and, as in
Proposition 1, $\,d=\min\{\dim\H_A,\dim\H_B\}\,$, $\,g(x)=(1+x)h_2\!\left(\frac{x}{1+x}\right)$.}\smallskip

A) \emph {If there exists an input state $\rho$ such that  $\,I_\mathrm{c}(\Phi,\rho)>0\,$, then}
\begin{equation}\label{DLB}
D_a(\Phi)\geq \frac{I_\mathrm{c}(\Phi,\rho)}{\log d}-\frac{1}{\log d}\;g\!\left(\frac{I_\mathrm{c}(\Phi,\rho)}{2\log d}\right)
\end{equation}
\emph{and}
\begin{equation}\label{EBLB-1}
D_{eb}(\Phi)\geq 2\frac{I_\mathrm{c}(\Phi,\rho)}{\log d}-\frac{2}{\log d}\;g\!\left(\frac{I_\mathrm{c}(\Phi,\rho)}{\log d}\right).
\end{equation}

B) \emph {If there is an input state $\rho$ such that $\,L(\Phi,\rho)\doteq H(\rho)-H(\widehat{\Phi}(\rho))>0\,$, then}
\begin{equation}\label{EBLB-2}
D_{eb}(\Phi)\geq 2\frac{L(\Phi,\rho)}{\log d}-\frac{2}{\log d}\;g\!\left(\frac{L(\Phi,\rho)}{\log d}\right).
\end{equation}

C) \emph {If  $\mathcal{H}_R\cong\mathcal{H}_A$ and $\;\omega$ is an arbitrary state in $\S(\H_{AR})$, then
\begin{equation}\label{EBLB-3}
D_{eb}(\Phi)\geq 2\frac{E_R(\Phi\otimes\id_R(\omega))}{\log d}-\frac{2}{\log d}\;g\!\left(\frac{E_R(\Phi\otimes\id_R(\omega))}{\log d}\right),
\end{equation}
where $E_R$ is the relative entropy of entanglement in $\S(\H_{BR})$.}\smallskip

D) \emph{If there exists an input  state $\rho$ such that $\,I_\mathrm{c}(\Phi,\rho)<0\,$ then}
\begin{equation}\label{ALB}
D_d(\Phi)\geq \frac{-I_\mathrm{c}(\Phi,\rho)}{\log d}-\frac{1}{\log d}\;g\!\left(\frac{-I_\mathrm{c}(\Phi,\rho)}{2\log d}\right).
\end{equation}

\emph{Each lower bound (\ref{DLB})-(\ref{ALB}) is tight for large $d$ in the sense of Def.\ref{tight-lb}.}
\end{property}
\smallskip

\begin{remark}\label{VLB-r}
Lower bounds (\ref{EBLB-1}) and (\ref{EBLB-2}) can be considered as computable weakened versions of lower bound (\ref{EBLB-3}), which is hard to compute in general (see Example 2 below where all these bounds are indicated explicitly).
\end{remark}\smallskip

\emph{Proof.} The inequalities (\ref{DLB}) and (\ref{ALB}) are obtained by applying part B of Lemma \ref{app} to continuity bound (\ref{CI-CB}).

To prove  (\ref{EBLB-1})-(\ref{EBLB-3}) note that
$$
D_{eb}(\Phi)\geq \sup_{\omega_{AR}}D_\mathrm{s}(\Phi\otimes\id_R(\omega_{AR})),
$$
where the supremum is over all states in $\S(\H_{AR})$ and $D_\mathrm{s}(\rho_{BR})$ is the distance between a state $\rho_{BR}$ and the set of all separable states in $\S(\H_{BR})$ (discussed in Section 3). By simple convexity reasoning the above supremum can be taken only over pure states
$\omega_{AR}$. Thus, lower bounds (\ref{EBLB-1})-(\ref{EBLB-3}) follow from Proposition \ref{DS}, since it is easily seen that
$$
I(R\rangle B)_{\Phi\otimes\id_R(\omega_{AR})}=I_\mathrm{c}(\Phi,\omega_{A})\quad \textrm{and}\quad I(B\rangle R)_{\Phi\otimes\id_R(\omega_{AR})}=L(\Phi,\omega_{A})
$$
for any pure state $\omega_{AR}$. \smallskip

The tightness of lower bounds (\ref{DLB}) and (\ref{ALB}) can be shown by using  the family of erasure channels (\ref{era-ch}).

It is known that the channel $\Phi_{\mathrm{p}}$ is degradable if $p\leq1/2$ and antidegradable if $p\geq1/2$ and that $I_\mathrm{c}(\Phi,\rho)=(1-2p)H(\rho)$ for all $p\in[0,1]$ \cite[Ch.10]{H-SCI}. So, if $\Phi=\Phi_{1/2-x}$ and $\rho$ is the chaotic state then the right hand side of (\ref{DLB}) is equal to
\begin{equation}\label{e-tmp}
2x-\frac{g(x)}{\log d},\quad\textrm{while}\quad D_a(\Phi_{1/2-x})\leq \|\Phi_{1/2-x}-\Phi_{1/2}\|_{\diamond}=2x.
\end{equation}
The tightness of (\ref{ALB}) is proved similarly by using the channel $\Phi_{1/2+x}$.\smallskip

The tightness of lower bounds (\ref{EBLB-1})-(\ref{EBLB-3}) follows from Example 1 below. $\square$

\smallskip
\textbf{Example 1.} If $\dim\H_A\leq\dim\H_B$, $\Phi=\id_A$ is the identity embedding of $\S(\H_A)$ into $\S(\H_B)$  and  $\rho$  is the chaotic  state in $\S(\H_A)$ then lower bound (\ref{DLB})  implies
\begin{equation*}
D_a(\id_A)\geq 1-\frac{g(1/2)}{\log d_A}\geq 1-\frac{1}{\log d_A},
\end{equation*}
where $d_A=\dim\H_A$, while (\ref{EBLB-1})-(\ref{EBLB-3}) give the same lower bound
\begin{equation*}
D_{eb}(\id_A)\geq 2-\frac{2g(1)}{\log d_A}\geq 2-\frac{2.8}{\log d_A}.
\end{equation*}
Thus, the radius of the maximal ball vicinity of $\id_A$ not containing antidegradable (respectively, entanglement-breaking)  channels is close to $1$ (respectively, to $2$) for large dimension $d_A$. On the other hand, it follows from (\ref{e-tmp}) with $x=1/2$ and the definition of the diamond norm that
$$
D_a(\id_A)\leq 1\quad\textrm{ and }\quad D_{eb}(\id_A)\leq 2.
$$
for any dimension $d_A$.\smallskip

\smallskip\textbf{Example 2.} If $\Phi_p$ is an erasure channel (\ref{era-ch}) then, evidently,
$I_\mathrm{c}(\Phi_p,\rho)\!=\!(1-2p)H(\rho)$ and $L(\Phi_p,\rho)=(1-p)H(\rho)-h_2(p)$ where $h_2$ is the binary entropy \cite[Ch.6]{H-SCI}. So, lower bounds (\ref{EBLB-1}) and (\ref{EBLB-2}) with the chaotic state $\shs\rho\shs$ imply respectively
\begin{equation*}
D_{eb}(\Phi_p)\geq 2(1-2p)-\frac{2}{\log d_A}g(1-2p)
\end{equation*}
and
\begin{equation*}
D_{eb}(\Phi_p)\geq 2(1-p)-\frac{2}{\log d_A}\left(h_2(p)+g\left((1-p)-\frac{h_2(p)}{\log d_A}\right)\right),
\end{equation*}
where it is assumed that $g(x)=0$ if $x<0$.\smallskip

Since $\Phi_p\otimes\id_R(\omega)=(1-p)\omega\oplus p|\varphi\rangle\langle\varphi|\otimes\Tr_A\omega$, where $\varphi$ is a unit vector in $\H_B\ominus\H_A$, by using basic properties of the relative entropy and convexity of $E_R$ one can show that $E_R(\Phi_p\otimes\id_R(\omega))=(1-p)E_R(\omega)$. Hence, lower bound (\ref{EBLB-3}) with maximally entangled state $\omega$ implies
\begin{equation*}
D_{eb}(\Phi_p)\geq 2(1-p)-\frac{2}{\log d_A}g(1-p).
\end{equation*}

Since $D_{eb}(\Phi_p)\leq\|\Phi_p-\Phi_1\|_{\diamond}=2(1-p)$, we see that
lower bounds (\ref{EBLB-2}) and (\ref{EBLB-3}) give tight lower estimates of $D_{eb}(\Phi_p)$ for large dimension $d_A$ (in contrast to  (\ref{EBLB-1})). We also see that lower bound (\ref{EBLB-3}) gives the sharpest estimate of $D_{eb}(\Phi_p)$ for all $p$ (as it was reasonable to expect). Unfortunately, application of this lower bound to arbitrary channel $\Phi$ is limited by the hard computability of $E_R$.

\section{Concluding remarks}

We have considered a nonstandard (inverse) application of tight continuity bounds for entropic quantities providing tight lower bounds on distance from a given quantum state (channel) to a certain class of states (channels). This approach is quite universal and can be used for different tasks. In particular, in view of Lemma \ref{app} the continuity bound (\ref{MI-CB}) and its extension to quantum conditional mutual information (see \cite{CHI}, Corollary 1) entail tight lower bounds for
\begin{itemize}
  \item  the $\|.\|_1$-distance between a given bipartite state $\rho_{AB}$ and  the set of all product states;
  \item  the $\|.\|_1$-distance between a given tripartite state $\rho_{ABC}$ and  the set of all short Markov chains (i.e. the states $\sigma_{ABC}$ such that $\sigma_{ABC}=\id_A\otimes\Phi(\sigma_{AB})$ for some channel $\Phi:B\rightarrow BC$ \cite{H&Co+}).
\end{itemize}

\bigskip

The authors are grateful to A.S.Holevo and A.Winter for useful discussion.
M.E.Shirokov acknowledges the support of the grant of Russian Science Foundation
(project No 14-21-00162).

\end{document}